\documentclass[10pt,journal]{IEEEtran}

\IEEEoverridecommandlockouts

\usepackage{amsmath,amsfonts,amssymb,color}
\usepackage{amsthm}
\usepackage{epsfig}
\usepackage{psfrag}
\usepackage{algorithm}
\usepackage{algpseudocode}
\usepackage{array}
\usepackage{epstopdf}
\usepackage{tikz, subcaption, cite}
\usepackage{multicol,multirow}

\newtheorem{problem}{Problem}
\newtheorem{theorem}{Theorem}

\newtheorem{lemma}{Lemma}
\newtheorem{remark}{Remark}
\newtheorem{definition}{Definition}
\newtheorem{proposition}{Proposition}
\newtheorem{example}{Example}
\newtheorem{assumption}{Assumption}

\begin{document}
\title{Control Policies for Recovery of Interdependent Systems After Disruptions}
\author{Hemant Gehlot, Shreyas Sundaram, and Satish V. Ukkusuri
\thanks{Hemant Gehlot and Satish V. Ukkusuri are with the Lyles School of Civil Engineering at Purdue University. Email: {\tt \{hgehlot,sukkusur\}@purdue.edu}. Shreyas Sundaram is with the School of Electrical and Computer Engineering at Purdue University. Email: {\tt sundara2@purdue.edu}. This research was supported by National Science Foundation award CMMI 1638311.}
}

\maketitle
\thispagestyle{empty}
\pagestyle{empty}
\begin{abstract}
We examine a control problem where the states of the components of a system deteriorate after a disruption, if they are not being repaired by an entity. There exist a set of dependencies in the form of precedence constraints between the components, captured by a directed acyclic graph (DAG). The objective of the entity is to maximize the number of components whose states are brought back to the fully repaired state within a given time. We prove that the general problem is NP-hard, and therefore we characterize near-optimal control policies for special instances of the problem. We show that when the deterioration rates are larger than or equal to the repair rates and the precedence constraints are given by a DAG, it is optimal to continue repairing a component until its state reaches the fully recovered state before switching to repair any other component. Under the aforementioned assumptions and when the deterioration and the repair rates are homogeneous across all the components, we prove that the control policy that targets the healthiest component at each time-step while respecting the precedence and time constraints fully repairs at least half the number of components that would be fully repaired by an optimal policy. Finally, we prove that when the repair rates are sufficiently larger than the deterioration rates, the precedence constraints are given by a set of disjoint trees that each contain at most $k$ nodes, and there is no time constraint, the policy that targets the component with the least value of health minus the deterioration rate at each time-step while respecting the precedence constraints fully repairs at least $1/k$ times the number of components that would be fully repaired by an optimal policy. 
\end{abstract}
\section{Introduction} \label{sec:intro}
We focus on a problem where multiple components of a system have been damaged after a disruption (e.g., a disaster, cyberphysical attack, etc.) and the states of the components continue to reduce if they are not being repaired, until they reach a value referred to as \textit{permanent failure}. There is an entity (or a controller) whose objective is to maximize a performance criteria or reward, e.g., maximizing the number of components whose states are brought back to their maximum value (known as the \textit{permanent repair} state) within a given time. The states of the components do not change once they reach permanent failure or permanent repair. There exists a set of \textit{dependencies} between the components that are represented by \textit{precedence} constraints  such that if there is a precedence constraint between an ordered pair of components $(i,j)$, then component $i$ needs to be permanently repaired before the entity can target component $j$. This problem has relevance in multiple applications, e.g., disaster recovery, protection of cyberphysical systems against attacks, and fire-fighting. For instance, infrastructure components face accelerated deterioration after disasters due to processes such as corrosion and flooded roads. If the components are not repaired in a timely manner, they can deteriorate to such a level that they become unusable and require full replacement, which is usually expensive \cite{chisolm2012impact}. Furthermore, there can be various dependencies between infrastructure components; for example, in order to bring back the functionality of water systems (such as irrigation pumps), it might be necessary to first repair some sections of the power network. 
Similarly, there might be a limited time before components (e.g., computing elements) of a cyberphysical system become fully compromised after an attack \cite{leversage2008estimating}. There might be dependencies or hierarchies between the components such that the components at the top of the hierarchy may be required to be repaired first in order to stop the spread of attack to other components \cite{la2015interdependent}. Thus, the protecting agency would have to make optimal recovery decisions if there is a repair budget (e.g., manpower and resources). 

This problem can be cast in the general framework of optimal control and scheduling of switched systems \cite{gorges2010optimal,bemporad2006logic}. In switched systems, there are multiple subsystems such that only one subsystem can be controlled at a time. In our problem, components correspond to the  subsystems of a switched system. Since the entity can only control one component at a time, our problem is analogous to optimal control and scheduling of discrete time linear switched systems \cite{gorges2010optimal}. One of the main complexities that arises in solving switched controlled systems is the combinatorially large number of feasible switching sequences \cite{bemporad2006logic}. Despite recent advances in the optimal control and scheduling of switched systems\cite{gorges2010optimal}, 
finding general theoretical results or frameworks for efficiently solving optimal control and scheduling problems for switched systems remains an active area of research \cite{zhu2015optimal}. 

At a high level, our problem also has analogies to the resource (e.g., time slot) allocation problem at a base-station to time-dependent flows \cite{eryilmaz2007fair}, optimal control of robotic systems \cite{smith2012persistent}, switched system control problems such as scheduling of thermostatically controlled loads \cite{nilsson2017class} and patient triage scheduling problems \cite{argon2008scheduling}. However, these studies do not consider precedence constraints. In addition, some of these studies \cite{eryilmaz2007fair,smith2012persistent} do not consider permanent failure of flows/components being served/targeted. 
Job scheduling with linear time deterioration and precedence constraints that minimize the total number of late jobs also have similarities to our problem \cite{gordon2008single}. However, job scheduling studies typically do not consider the notion of permanent failure, as jobs are completely processed even if their completion times exceed due dates. In addition, jobs are considered to be late if their \textit{completion} times exceed the corresponding due dates; however, in our setting, a repair agency should \textit{start} targeting a component before it fails. Scheduling of real-time tasks \cite{zhang2009schedulability} also has some high level analogies to our work. However, these studies focus on tasks that sporadically become available for processing, in contrast to the components in our problem that are available for repair starting at the same time (i.e., immediately after a disruption), subject to precedence constraints.

Under the setting described before, we focus on characterizing (near-) optimal switching policies for maximizing the number of components that are
permanently repaired. The paper \cite{hgehlot2019ACC} characterized optimal policies for this problem assuming homogeneous rates of repair and deterioration, and \cite{hgehlot_tac} studied the problem for the case when there are heterogeneous rates of repair and deterioration. However, these studies do not consider precedence constraints among various components and also do not constrain the maximum time that is available to repair the components. In the conference version of this paper \cite{gehlot2019approximation}, precedence constraints were considered among various components but there was no constraint on the time that is available for repair, and no discussion on the complexity of the general problem. This paper significantly expands on the conference paper by addressing those gaps. 

\subsection*{Our contributions}
First, we prove that when the deterioration rates are larger than or equal to the repair rates, and the precedence constraints are given by a directed acyclic graph (DAG), the optimal switching policy is to permanently repair a component that is targeted before switching to another component, while respecting the precedence and time constraints. Second, we prove that under the aforementioned assumptions and when the deterioration and the repair rates are homogeneous across all the components, the switching policy that targets the healthiest component at each time-step while respecting the precedence and time constraints is $1/2$-optimal.\footnote{For $\alpha\in (0,1]$, a policy is said to be $\alpha$-optimal if the reward obtained by that policy is at least $\alpha$ times the reward obtained by an optimal policy.} We also prove that the aforementioned policy is optimal when time is constrained but when there are no precedence constraints. Third, we prove that when the repair rates are sufficiently larger than the deterioration rates, the precedence constraints are given by a set of disjoint trees that each contain at most $k$ nodes, and there is no time constraint, the policy that targets the component with the least health value minus the deterioration rate at each time-step while respecting the precedence constraints is $1/k$-optimal.\footnote{The paper \cite{gehlot2019approximation} characterizes a near-optimal policy with the same factor $(1/k)$ under the aforementioned assumptions where the root nodes of the trees are first targeted and other nodes are only targeted when it is not possible to target any root node; the policy that is characterized in this paper does not require that condition.} 
Finally, we prove that the general problem is NP-hard (even if there is no time constraint). 

In the next section, we formally define the problem that we focus in this paper. After that, we divide the characterization of sequencing policies into two cases, corresponding to the relative sizes of the rates of deterioration and repair. 
\section{Problem Statement}
There are $N(\ge 2)$ components indexed by the set $\mathcal{V}=\{1,\ldots,N\}$. A component could represent damaged infrastructure in an area, an infected server, etc., depending on the context. We refer to different components as \emph{nodes}. There is a repair agency (also referred to as an  \emph{entity}), which is responsible for repairing the nodes. We index time-steps by the variable $t\in \mathbb{N} = \{0, 1, 2, \ldots\}$, capturing the resolution at which the entity decides to repair nodes. Let $T\in\mathbb{N} \cup \{\infty\}$ be the largest time-step that is available for repairing nodes (when $T= \infty$ we say that there is no time constraint). The control action of the entity at time-step $t$ is represented by $u_t\in \mathcal{V}$, i.e., the entity targets one node to repair at each time-step.\footnote{We keep the analysis of the problem where the entity can simultaneously target multiple nodes for future work.} 
Each node $j\in \mathcal{V}$ has an associated health value $v_t^j \in [0,1]$ at time-step $t$. We denote the initial health value of each node $j$ by $v_0^j\in (0,1)$. 
If a node $j$ is being repaired at a time-step then the health value of that node increases by $\Delta_{inc}^j \in (0,1)$. A node $j$ is considered to be \textbf{permanently repaired} at time-step $t$ if $v^j_t=1$ and $v^j_{t-1}<1$. If a node $j$ is not being repaired at a time-step, then the health value of node $j$ decreases by $\Delta_{dec}^j \in (0,1)$. A node $j$ is considered to be \textbf{permanently failed} at time-step $t$ if $v^j_t=0$ and $v^j_{t-1}>0$. Note that the health value of a node does not change further once it gets permanently repaired or permanently failed. Thus, for all $j\in \{1,\ldots,N\}$, the health value of node $j$ dynamically changes depending on the control action taken by the entity as follows:
\begin{equation*}
v^j_{t+1} = \begin{cases} 1 &\text{if } v^j_t=1, \\
0 &\text{if } v^j_t=0, \\
\min(1,v^j_t + \Delta_{inc}^j) & \hbox{if } u_t = j \hspace{1mm}\text{and}\hspace{1mm} v^j_t\in(0,1), \\
\max(0, v^j_t-\Delta_{dec}^j) & \hbox{if } u_t \neq j \hspace{1mm}\text{and}\hspace{1mm} v^j_t\in(0,1).\end{cases}
\end{equation*}

There is a set of dependency constraints between different nodes,  represented by a set $\mathcal{E}=\{(j,k)|j,k\in \mathcal{V}\}$ consisting of edges between various nodes. An edge $(j,k)\in \mathcal{E}$ starting from node $j$ and ending at node $k$ represents a precedence constraint that node $j$ needs to be permanently repaired (i.e., to full health) before the entity can start targeting node $k$; $j$ is an \textit{in-neighbor} of $k$. The \textit{in-neighbor set} of a node $k\in \mathcal{V}$ is the set $\mathcal{W}_k=\{j\in\mathcal{V}|(j,k)\in \mathcal{E}\}$. We assume that the precedence constraints form a directed acyclic graph (DAG) denoted by $G=\{\mathcal{V},\mathcal{E}\}$; such graphs are a common form for representing general precedence constraints in other problems such as job scheduling \cite{gordon2008single}. 

We now define the \textit{feasible control set} as follows.
\begin{definition}
Let the health values of all the nodes at time-step $t$ be given by $v_t=\{v_t^1,\ldots,v_t^N\}$. Then, the feasible control set at time-step $t$ is the set $\mathcal{U}(v_t)\subseteq \mathcal{V}$ containing the nodes whose in-neighbors are all in the permanently repair state at time-step $t$, i.e., $\mathcal{U}(v_t)=\{k\in \mathcal{V}|v^j_t=1, \forall j \in \mathcal{W}_k\}$. 
\end{definition}

We now present the definition of a \textit{control sequence that respects the precedence constraints} as follows.
\begin{definition}
A control sequence $\bar{u}_{0:T}=\{\bar{u}_{0},\bar{u}_{1},\ldots,\bar{u}_{T}\}$ is said to respect the precedence constraints if for all $t\in[0,T]$, $\bar{u}_t\in \mathcal{U}(v_t)$.
\end{definition}

We define the reward function as follows.
\begin{definition} \label{defn:reward}
Given a set of initial health values $\{v_0^j\}$ and a control sequence $\bar{u}_{0:T}$ that respects the precedence constraints, the reward $J(v_0, \bar{u}_{0:T})$ is defined as the number of nodes that get permanently repaired through that sequence. More formally, $J(v_0, \bar{u}_{0:T})=|\{j\in \mathcal{V}\hspace{1mm}|\hspace{1mm}\exists \hspace{1mm} t \text{ s.t. } 0\le t\le T \text{ and }v^j_{t+1}=1\}|$.  
\end{definition}

Based on the above definitions, the following problem is the focus of this paper.  
\begin{problem}[Optimal Control Sequencing over a DAG] \label{problem} 
$ $\newline Consider a directed acyclic graph $G=\{\mathcal{V},\mathcal{E}\}$ consisting of $N(\ge 2)$ nodes with initial health values $\{v_0^j\}$, along with repair and deterioration rates $\{\Delta_{inc}^j\}$ and $\{\Delta_{dec}^j\}$, respectively. Given a time constraint $T\in \mathbb{N} \cup \{\infty\}$, find a control sequence $u^{\ast}_{0:T}$ that respects the precedence constraints and maximizes the reward $J(v_0, u^{\ast}_{0:T})$.
\end{problem}

For our purposes, we also define the concept of a \textbf{jump}. 
\begin{definition}
 If the entity starts targeting a node before permanently repairing the node it targeted in the last time-step, then the entity is considered to have jumped at that time-step. Mathematically, if $u_{t-1}=j$, $v_{t}^j<1$ and $u_{t} \neq j$, then a jump is said to have been made by the entity at time-step $t$. A control sequence that does not contain any jumps is called a \textbf{non-jumping sequence}.
\end{definition}

In this paper, we will be obtaining sequences that are generated by state feedback policies, i.e., we will be providing a mapping $\mu: [0,1]^N \rightarrow \mathcal{V}$, such that $\bar{u}_t = \mu(v_t)$ for all $t\in[0,T]$. Thus, we will use the same terminology defined above (for sequences) for policies (e.g., a policy respecting the precedence constraints, a non-jumping policy, etc.). Given a policy $\mu$ that respects the precedence constraints, we denote the reward (from Definition \ref{defn:reward}) by $J_{\mu}(v_0)$, with the time-constraint being implicit.

We now define an $\alpha$-optimal policy as follows.  
\begin{definition}
Let $\mu^*$ be an optimal policy for Problem \ref{problem}. For $\alpha\in(0,1]$, a policy $\mu$ (that respects the precedence and time constraints) is said to be $\alpha$-optimal if $J_{\mu}(v_0)\ge \alpha J_{\mu^*}(v_0), \forall v_0\in [0,1]^N$. 
\end{definition}

In the next section, we start the analysis of the problem for $\Delta_{dec}^j\ge \Delta_{inc}^j, \quad \forall j \in \{1,\ldots,N\}$.

\section{Control Sequences for $\Delta_{\MakeLowercase{dec}}^{\MakeLowercase{j}}\ge \Delta_{\MakeLowercase{inc}}^{\MakeLowercase{j}}, \quad \forall \MakeLowercase{j} \in \{1,\ldots,N\}$}
In this section, we start by showing the optimality of non-jumping sequences when $\Delta_{dec}^j\ge \Delta_{inc}^j, \quad \forall j \in \{1,\ldots,N\}$.
\subsection{Optimality of non-jumping sequences}

To show that non-jumping policies are optimal when $\Delta_{dec}^j\ge \Delta_{inc}^j, \forall j\in \{1,\ldots,N\}$, we first analyze the properties of sequences containing at most one jump and later generalize to sequences containing an arbitrary number of jumps. We start with the following result.

\begin{figure}[ht]
	\begin{center}
		\includegraphics[scale=0.65]{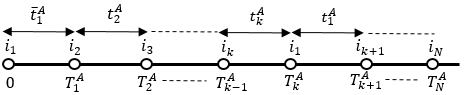}
	\end{center}	
	\caption{Sequence $A$ with a single jump.}
	\label{seqA}
\end{figure}

\begin{figure}[ht]
	\begin{center}
		\includegraphics[scale=0.65]{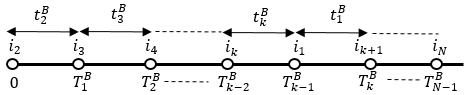} 
	\end{center}	
	\caption{Non-jumping sequence $B$.}
	\label{seqB}
\end{figure}

\begin{lemma} \label{lem:newsinglejumplem}
Let there be $N(\ge 2) $ nodes that have a set of precedence constraints given by a DAG $G=\{\mathcal{V},\mathcal{E}\}$, $\Delta_{dec}^j\ge \Delta_{inc}^j, \forall j\in \{1,\ldots,N\}$ and $T\in \mathbb{N} \cup \{\infty\}$. Consider a control sequence $A$ that respects the precedence and time constraints and targets $N$ nodes as shown in Figure \ref{seqA}. Suppose sequence $A$ permanently repairs all the nodes and contains exactly one jump, where the entity partially repairs node $i_1$ before moving to node $i_2$ at time-step $\bar{t}_{1}^A$. Sequence $A$ then permanently repairs nodes $i_2, i_3, \ldots, i_k$, before returning to node $i_1$ and permanently repairing it. Then, there exists a non-jumping control sequence $B$ as shown in Figure \ref{seqB} that targets nodes in the order $\{i_2, i_3, \ldots, i_k, i_1, i_{k+1}, \ldots, i_N\}$ (which respects the precedence and time constraints) and permanently repairs all the nodes. Furthermore, if $t_j^A$ (resp. $t_j^B$) is the number of time-steps taken to permanently repair node $i_j$ in sequence $A$ (resp. sequence $B$), then the following holds true:
\begin{align}
t_j^A &\geq t_j^B+\left(2^{j-2}\right)\overline{t}_1^A \hspace{3mm}\forall j \in \{2,\ldots,k\}, \label{claim1} \\
t_1^A &\geq t_1^B+(2^{k-1}-2)\overline{t}_1^A, \label{claim2} \\
t_j^A &\geq t_j^B+\left(2^{j-1}-2^{j-k}\right)\overline{t}_1^A \hspace{3mm}\forall j \in \{k+1,\ldots,N\}. \label{claim3}
\end{align}
\end{lemma}

\begin{IEEEproof}
Since node $i_1$ is partially repaired by the entity before it targets the nodes $i_2, \ldots,i_k$ in sequence $A$, there cannot be an edge that starts from $i_1$ and ends at a node in the set $\{i_2, \ldots,i_k\}$. Thus, if sequence $A$ follows the precedence constraints, the non-jumping sequence $B$ also respects the precedence constraints. 
We can now leverage the analysis in \cite{hgehlot_tac}, which proved that \eqref{claim1}-\eqref{claim3} hold and each node in sequence $B$ is targeted at an earlier time-step than in sequence $A$. Therefore, sequence $B$ permanently repairs all the nodes that are permanently repaired by sequence $A$ while respecting the precedence and time constraints.
Thus, the result follows. 
\end{IEEEproof}
The above result considered sequences containing exactly one jump. This leads us to the following key result pertaining to the optimal control policy when $\Delta_{dec}^j\ge \Delta_{inc}^j, \forall j$.

\begin{theorem} \label{thm:nonjumping_optimal_notallfixed}
	Let there be $N (\geq 2)$ nodes that have a set of precedence constraints given by a DAG $G=\{\mathcal{V},\mathcal{E}\}$, $\Delta_{dec}^j\ge \Delta_{inc}^j, \forall j\in \{1,\ldots,N\}$ and $T\in \mathbb{N} \cup \{\infty\}$. Suppose there is a sequence $A$ with one or more jumps that respects the precedence and time constraints, and permanently repairs $x(\le N)$ nodes. Then, there exists a non-jumping sequence that respects the precedence and time constraints, and permanently repairs $x$ nodes. Thus, non-jumping sequences are optimal when $\Delta_{dec}^j\ge \Delta_{inc}^j, \forall j\in \{1,\ldots,N\}$.
\end{theorem}

The proof of the above result follows immediately by applying Lemma \ref{lem:newsinglejumplem} on sequence $A$ to obtain a sequence that has one less jump than $A$, and repeating this procedure on the obtained sequences to get a non-jumping sequence. 

Since non-jumping sequences are optimal by the above result, we will only focus on non-jumping sequences when $\Delta_{dec}^j\ge \Delta_{inc}^j, \forall j\in \{1,\ldots,N\}$. We now present the following NP-hardness result, whose proof is provided in the appendix. 
\begin{theorem} \label{thm:NP-hard}
Problem \ref{problem} is NP-hard.
\end{theorem}

Since Problem \ref{problem} is NP-hard, it is unlikely that the optimal solution can be efficiently computed for all instances of the problem \cite{cormen2009introduction}. Thus, we will characterize optimal or near-optimal policies for special instances of the problem. 

\subsection{Near-optimal policy}
In this section, we characterize a near-optimal policy for the case when $\Delta_{dec}^j\ge \Delta_{inc}^j, \forall j\in \{1,\ldots,N\}$. 
We will make the following assumption in this section.
\begin{assumption} \label{assum:no_partaway_step}
For all  $j\in\{1,\ldots,N\}$, suppose $\Delta_{inc}^j=\Delta_{inc}$, $\Delta_{dec}^j=\Delta_{dec}$, and $\Delta_{dec}\ge \Delta_{inc}$. Also, suppose there exists a positive integer $n$ such that $\Delta_{dec} = n\Delta_{inc}$, and for each node $j\in\{1,\ldots,N\}$, there exists a positive integer $m_j$ such that $1-v^{j}_0 =m_j \Delta_{inc}$.
\end{assumption}
Note that the conditions $\Delta_{dec} = n\Delta_{inc}$ and $1-v^{j}_0 =m_j \Delta_{inc},\hspace{2mm} \forall j \in \{1,\ldots,N\}$, in the assumption ensure that no node gets permanently repaired partway through a time-step.

We now start with the following result.

\begin{lemma} \label{lem:onelessnode}
Let there be $N(\geq2)$ nodes with precedence constraints given by a DAG $G=\{\mathcal{V},\mathcal{E}\}$. Suppose Assumption \ref{assum:no_partaway_step} holds and $T\in \mathbb{N} \cup \{\infty\}$. Consider a non-jumping sequence $A$ that respects the precedence and time constraints, and permanently repairs $x(\le N)$ nodes. Suppose there is a node $h$ from the set of $N$ nodes that has larger initial health value than the first node of sequence $A$ and the precedence constraints allow node $h$ to be permanently repaired before the first node of sequence $A$. Consider a non-jumping sequence $B$ where node $h$ is first targeted and then the remaining targeted nodes of sequence $A$ are targeted after node $h$. Then, at least $x-1$ nodes are permanently repaired in sequence $B$ while respecting the precedence and time constraints.
\end{lemma}

\begin{IEEEproof}
Let $x$ be the number of nodes that are permanently repaired in sequence $A$. Denote the order of nodes in the sequence $A$ as $\{i_1,\ldots,i_x\}$. The proof for the case when $x=1$ is trivial because $x-1=0$; we thus focus on the case when $x\ge 2$. Denote the ordering of the first $x-1$ nodes in sequence $B$ as $\{i'_1,\ldots,i'_{x-1}\}$, where $i'_1=h$. 
Note that the proof when $x=2$ is also trivial because node $i'_1=h$ in the non-jumping sequence $B$ is permanently repaired since the time taken to permanently repair node $h$ in sequence $B$ is less than the time taken to permanently repair the nodes in sequence $A$ (because node $h$ has larger initial health than node $i_1$). Therefore, the non-trivial cases are when $x\ge 3$. 

We first argue that sequence $B$ permanently repairs at least $x-1$ nodes, regardless of the time it takes to repair the nodes. We prove the result by contradiction. Suppose sequence $B$ permanently repairs less than $x-1$ nodes. Let $i'_k$ be the first node that permanently fails in sequence $B$, where $2\le k\le x-1$. We first argue that either $i'_k=i_{k-1}$ or $i'_k=i_{k}$. If node $h$ does not appear in sequence $A$, then the first $x-1$ nodes of sequence $B$ are ordered as follows $\{h,i_1,\ldots,i_{x-2}\}$ and thus $i'_k=i_{k-1}$. On the other hand, if node $h$ is permanently repaired in sequence $A$, let the position of node $h$ in sequence $A$ be $z$, where $z\in \{2,\ldots,x\}$. Then, $i'_1=h$, $i'_{j}=i_{j-1},\quad\forall j\in \{2,\ldots,z\}$ and $i'_{j}=i_{j},\quad \forall j\in \{z+1,\ldots,x-1\}$. Thus, if the first node that fails in sequence $B$ (i.e., node $i'_k$) is such that $k\le z$, then $i'_k=i_{k-1}$; otherwise $i'_k=i_k$. We now consider the cases when $i'_k=i_{k-1}$ and $i'_k=i_k$ one by one. 

Suppose $i'_k=i_{k-1}$. If node $i_{k-1}$ fails in sequence $B$, then we show that node $i_{k+1}$ must permanently fail in sequence $A$ (where $2\le k\le x-1$). Consider $k=2$. Let $t_h=\frac{1-v_0^h}{\Delta_{inc}}$ be the number of time-steps taken to permanently repair node $h$ from its initial health to the full health. If node $i_1$ permanently fails after permanently repairing node $h$ in sequence $B$, then $v_0^{i_1}\le \Delta_{dec}t_h$. That is,
\begin{equation}
    1-v_0^{i_1}\ge 1-\Delta_{dec}t_h. \label{eq:3node_firstnode}
\end{equation}
Denote $t_j=\frac{1-v_0^{i_j}}{\Delta_{inc}}$ as the number of time-steps taken to permanently repair node $i_j$ from its initial health to full health. Then, 
\begin{equation}
    \Delta_{dec}t_1=\Delta_{dec}\left(\frac{1-v_0^{i_1}}{\Delta_{inc}}\right)=n\left(1-v_0^{i_1}\right)\ge \left(1-v_0^{i_1}\right), \label{eq:3node_firstnode_2}
\end{equation}
as $n\ge 1$ and $v_0^{i_1}<1$. Therefore, 
\begin{equation}
    \Delta_{dec}t_1\ge 1-\Delta_{dec}t_h>1-\Delta_{dec}t_1, \label{eq:3node_firstnode_3}
\end{equation}
where the first inequality from the left comes by conditions \eqref{eq:3node_firstnode}-\eqref{eq:3node_firstnode_2} and the second inequality comes from the fact that $t_h<t_1$ (as node $h$ has larger initial health than node $i_1$). Note that in sequence $A$, it takes $t_1$ time-steps to repair node $i_1$, and  $nt_1+t_2$ time-steps to repair node $i_2$ (it takes $nt_1$ time-steps to repair the health that is lost by node $i_2$ due to deterioration and it takes $t_2$ time-steps to repair the difference in the initial health of $i_2$ and full health). By \eqref{eq:3node_firstnode_3}, the total reduction in the health of node $i_3$ after $(1+n)t_1+t_2$ time-steps in sequence $A$ satisfies
\begin{align*}
    \Delta_{dec}\left((1+n)t_1+t_2\right)  &=\Delta_{dec}\left(nt_1+t_2\right)+\Delta_{dec} t_1 \\
    &>\Delta_{dec}\left(nt_1+t_2\right)+1-\Delta_{dec}t_1\\
    &>1,
\end{align*}
as $n\ge 1$ and $t_1,t_2>0$.  Thus, node $i_3$ permanently fails in sequence $A$ by the time it is reached, contradicting the fact that $A$ permanently repairs $x (\ge 3)$ nodes. 

Now consider the case when $k=3$. If node $i_2$ permanently fails in sequence $B$, then $v_0^{i_2}\le \Delta_{dec}((1+n)t_h+t_1)$. Following the same arguments as before, we get
\begin{equation}
    \Delta_{dec}t_2>1-\Delta_{dec}\left((2+n)t_1\right). \label{eq:jequalto2_first}
\end{equation}
As before, the total number of time-steps required to permanently repair nodes $i_1, i_2$, and $i_3$ in sequence $A$ is equal to
$(1+n)^2t_1+(1+n)t_2+t_3$.
Thus, the total reduction in the health of node $i_4$ by the time it is reached in sequence $A$ satisfies
\begin{equation*}
    \Delta_{dec}\left((1+n)^2t_1+(1+n)t_2+t_3\right) >1,
\end{equation*}
by \eqref{eq:jequalto2_first}, $n\ge 1$ and $t_1,t_2,t_3>0$.  Thus, $i_4$ permanently fails by the time it is reached in sequence $A$, leading to a contradiction.

We can repeat the above arguments to show that if node $i_{k-1}$ permanently fails in sequence $B$, where $k>3$, then
\begin{multline}
    \Delta_{dec}t_{k-1}>1-\Delta_{dec}((2+n)(1+n)^{k-3}t_1 + \\ (1+n)^{k-4}t_2+ \ldots+(1+n)^0t_{k-2}). \label{eq:generalnode_first}
\end{multline} 
Therefore, the total reduction in the health of node $i_{k+1}$ after $(1+n)^{k-1}t_1+(1+n)^{k-2}t_2+\ldots+(1+n)^{0}t_{k}$ time-steps in sequence $A$ is 
\begin{align}
    &\Delta_{dec}\left((1+n)^{k-1}t_1+\ldots+(1+n)^{1}t_{k-1}+(1+n)^{0}t_{k}\right) =\nonumber \\
    &\Delta_{dec}\left((1+n)^{k-1}t_1+\ldots+nt_{k-1}+(1+n)^{0}t_{k}\right)+\Delta_{dec}t_{k-1}. \label{eq:totaldecrease_hnotfixed}
\end{align}
Thus, for $k> 3$, we get
\begin{align*}
   \Delta_{dec}\left((1+n)^{k-1}t_1+\ldots+(1+n)^1t_{k-1}+(1+n)^{0}t_{k}\right)>1
\end{align*}
by \eqref{eq:generalnode_first}, \eqref{eq:totaldecrease_hnotfixed}, $n\ge 1$ and $t_j>0 \hspace{1mm} \forall j$. Since, the total reduction in the health value of node $i_{k+1}$ before the entity starts targeting it is larger than one, it is not possible to permanently repair node $i_{k+1}$ in sequence $A$, leading to a contradiction.

We now consider the case when $i'_k=i_k$ (along with $x\ge 3$). We prove that if node $i'_{k}=i_{k}$ permanently fails in sequence $B$, then it is not possible to permanently repair node $i'_{k+1}=i_{k+1}$ in sequence $A$. Recall that the case $i'_k=i_k$ happens when node $h$ is permanently repaired in sequence $A$ and $k>z$. Since $z\ge 2$, we have $k>2$. Consider $k=3$. Then, $z=2$ (i.e., $i_2=h$) because $z<k$ and $z \ge 2$. If node $i'_3=i_3$ permanently fails in sequence $B$, then
\begin{equation*}
    v_0^{i_3}\le \Delta_{dec}((1+n)t_h+t_1)<\Delta_{dec}\left((2+n)t_1\right),
\end{equation*}
or equivalently,
\begin{equation}
    \Delta_{dec}t_3>1-\Delta_{dec}\left((2+n)t_1\right). \label{eq:nodehinseqA_k=3}
\end{equation} 
Therefore, the total reduction in the health of node $i_4$ after $(1+n)^2t_1+(1+n)t_2+t_3$ time-steps in $A$ satisfies 
\begin{align*}
    &\Delta_{dec}\left((1+n)^2t_1+(1+n)t_2+t_3\right)\\
    &=\Delta_{dec}\left((1+n)^2t_1+(1+n)t_2\right)+\Delta_{dec}t_3>1,
\end{align*}
by \eqref{eq:nodehinseqA_k=3}, $n\ge 1$ and $t_1,t_2>0$.  Thus, node $i_4$ permanently fails in sequence $A$, contradicting the fact that $A$ permanently repairs $x$ nodes.

Similarly, it can be shown that if node $i'_{k}=i_{k}$ permanently fails in sequence $B$, where $k>3$, then
\begin{multline}
    \Delta_{dec}t_{k}>1-\Delta_{dec}((2+n)(1+n)^{k-3}t_1 + \\ (1+n)^{k-4}t_2+\ldots+(1+n)^{k-1-z}t_{z-1}+\\ (1+n)^{k-2-z}t_{z+1}+\ldots+(1+n)^0t_{k-1}). \label{eq:nodehinseqA_k>3}
\end{multline} 
Therefore, the total reduction in the health of node $i_{k+1}$ (where $k> 3$) after $(1+n)^{k-1}t_1+(1+n)^{k-2}t_2+\ldots+(1+n)^{0}t_{k}$ time-steps in sequence $A$ is 
\begin{align*}
    &\Delta_{dec}\left((1+n)^{k-1}t_1+\ldots+(1+n)^{1}t_{k-1}+(1+n)^{0}t_{k}\right) \\
    &=\Delta_{dec}\left((1+n)^{k-1}t_1+\ldots+(1+n)^{1}t_{k-1}\right)+\Delta_{dec}t_k>1, 
\end{align*}
by \eqref{eq:nodehinseqA_k>3}, $n\ge 1$ and $t_j>0 \hspace{1mm} \forall j$. Therefore, it would not be possible to permanently repair $x$ nodes in sequence $A$, leading to a contradiction. 

We will now show that sequence $B$ permanently repairs $x-1$ nodes in less time than the time taken to permanently repair $x$ nodes in sequence $A$. Suppose node $h$ is not permanently repaired in sequence $A$. Consider $x=3$. Then, the time taken to permanently repair the first two nodes, i.e., nodes $h$ and $i_1$, in sequence $B$ is equal to $(1+n)t_h+t_1$ and the time taken to permanently repair three nodes in sequence $A$ is equal to $(1+n)^2 t_1+(1+n)t_2+t_3$. Note that $(1+n)t_h+t_1 <(2+n)t_1 <(1+n)^2 t_1+(1+n)t_2+t_3$ because $t_h<t_1$, $n\ge 1$ and $t_1,t_2,t_3>0$. Consider the case when $x>3$. Then, the time taken to permanently repair $x-1$ nodes in sequence $B$ is equal to
 \begin{equation}
      (1+n)^{x-2}t_h+(1+n)^{x-3}t_1+\ldots+(1+n)^0 t_{x-2}, \label{eq:time_seqB_hnotincluded}
  \end{equation}
  and the time taken to permanently repair $x$ nodes in sequence $A$ is equal to 
  \begin{equation}
     (1+n)^{x-1}t_1+(1+n)^{x-2}t_2+\ldots+(1+n)^0t_x. \label{eq:time_seqA_hnotincluded}
  \end{equation}
  Then, $(1+n)^{x-2}t_h+(1+n)^{x-3}t_1+(1+n)^{x-4}t_2+\ldots+(1+n)^0 t_{x-2}<\left((2+n)(1+n)^{x-3}\right)t_1+(1+n)^{x-4}t_2+\ldots+(1+n)^0 t_{x-2}<(1+n)^{x-1}t_1+(1+n)^{x-2}t_2+\ldots+(1+n)^0t_x$ as $t_h<t_1$, $n\ge 1$ and $t_j>0, \forall j\in \{1,\ldots,x\}$. 

Now consider the case when node $h$ is permanently repaired in sequence $A$. When $z\ge x-1$ (and $x\ge 3$), the time taken to permanently repair $x-1$ nodes in sequence $B$ and the time taken to permanently repair $x$ nodes in sequence $A$ are given by expressions \eqref{eq:time_seqB_hnotincluded}
 and \eqref{eq:time_seqA_hnotincluded}, respectively, and therefore the proof proceeds in the same way as when node $h$ is not permanently repaired in sequence $A$. Thus, the proof for $x=3$ follows in exactly the same way as before because $z \ge 2$.
 We now focus on the case when $z\le x-2$ and $x>3$. Then, the total time taken to permanently repair $x-1$ nodes in sequence $B$ is equal to
 \begin{multline}
   (1+n)^{x-2}t_h+(1+n)^{x-3}t_1+\ldots+(1+n)^{x-1-z}t_{z-1}+\\(1+n)^{x-2-z}t_{z+1}+\ldots+(1+n)^0 t_{x-1}, \label{eq:time_seqB_hincluded}
 \end{multline}
and the time taken to permanently repair $x$ nodes in sequence $A$ is given by the expression \eqref{eq:time_seqA_hnotincluded}. 
Note that the value of the expression \eqref{eq:time_seqB_hincluded} is less than the value of the expression \eqref{eq:time_seqA_hnotincluded} because $t_h<t_1$, $n\ge 1$ and $t_j>0, \forall j \in \{1,\ldots,x\}$. Therefore, if sequence $A$ permanently repairs $x$ nodes by time-step $T$, then sequence $B$ must permanently repair at least $x-1$ nodes by time-step $T$. Thus, the result follows.
\end{IEEEproof}
Using the above result, we now propose a near-optimal policy for a subclass of Problem \ref{problem}.
\begin{theorem} \label{thm:2-approx}
Let there be $N(\geq2)$ nodes with precedence constraints given by a DAG $G=\{\mathcal{V},\mathcal{E}\}$. Suppose Assumption \ref{assum:no_partaway_step} holds and $T\in \mathbb{N} \cup \{\infty\}$. Then, the policy that targets the healthiest node at each time-step while respecting the precedence and time constraints is $1/2$-optimal.
\end{theorem}
\begin{IEEEproof}
Let $A$ be the optimal (non-jumping) sequence. Then, we go to the first time-step $T'(\le T)$ in sequence $A$ at which the healthiest available node (i.e., the healthiest node that can be targeted at time-step $T'$ without violating the precedence constraints) is not targeted. Denote the portion of sequence $A$ from time-step $0$ to time-step $T'-1$ as $A'$ and the portion of sequence from time-step $T'$ onwards as $A''$. We modify the sequence $A''$ by Lemma \ref{lem:onelessnode} to generate sequence $A^*$ such that the healthiest available node (while respecting the precedence and time constraints) is targeted at time-step $T'$ in $A^*$. Let $y$ be the number of nodes that are permanently repaired by sequence $A''$. Then, at least $y-1$ nodes are permanently repaired in sequence $A^*$ (while respecting the precedence and time constraints) by Lemma \ref{lem:onelessnode}. Then, a sequence $B$ is generated by concatenating $A'$ with $A^*$. After this, we go to the next time-step in sequence $B$ at which the healthiest node (while respecting the precedence and time constraints) is not targeted and repeat this procedure. We iteratively repeat this procedure, so that we finally get a sequence in which the healthiest node (while respecting the precedence and time constraints) is targeted at each time-step. The number of nodes that are permanently repaired in the final sequence will be at least half of the number of nodes that are permanently repaired in sequence $A$ because 1) at each iteration of this procedure we move one node across the given sequence and the number of nodes that are permanently repaired in the modified sequence reduces at most by one, and 2) in the last iteration of this procedure where there is only one node, there is no reduction in the number of permanently repaired nodes because if the last node in the given sequence can be permanently repaired then a healthier node that replaces it can also be permanently repaired. 
\end{IEEEproof}

We now show that the factor of $1/2$ in the above result is sharp, in that there exist problem instances where targeting the healthiest node at each time-step repairs only half as many nodes as an optimal policy.   

\begin{figure}[ht]
	\begin{center}
		\includegraphics[scale=0.3]{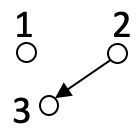}
	\end{center}	
	\caption{Graph for illustrating optimal policies in the presence of precedence constraints.}
	\label{fig:counterexample}
\end{figure}

\begin{example} \label{ex:counterex}
Consider a graph consisting of three nodes as shown in Figure \ref{fig:counterexample}. Let the initial health values of the nodes 1, 2 and 3 be 0.6, 0.3 and 0.8, respectively. The deterioration and repair rates are homogeneous across all the nodes and both are equal to 0.1. Suppose $T=\infty$. The healthiest node that can be targeted in the first time-step is node 1 (as we need to permanently repair node 2 before targeting node 3). If node 1 is first permanently repaired then node 2 fails by the time the entity reaches it and we cannot permanently repair any more nodes. However, if node 2 is first permanently repaired then node 3 can also be permanently repaired. Note that although the policy that targets the healthiest node at each time-step (while respecting the precedence and time constraints) is not optimal in this example, it is indeed $1/2$-optimal as proved in Theorem \ref{thm:2-approx}.
\end{example}

\begin{remark}
In this paper, we consider all the nodes to be equally important. However, one could also consider weights on the nodes to signify their relative importance, and the objective of Problem \ref{problem} can be changed to maximize the total weight of the nodes that are permanently repaired. Then, it is easy to argue that the policy characterized in Theorem \ref{thm:2-approx} would be $\frac{w_{min}}{2w_{max}}$-optimal, where $w_{max}$ and $w_{min}$ are the maximum and minimum weights of nodes, respectively.
\end{remark}

Theorem \ref{thm:2-approx} shows that the feedback policy that targets the healthiest node at each time-step while respecting the precedence and time constraints is $1/2$-optimal. In the next section, we show that this policy is, in fact, optimal for special instances of the problem.
\subsection{Optimal sequencing}
The paper \cite{hgehlot_tac} showed that the policy that targets the healthiest node at each time-step is optimal when all the conditions of Theorem \ref{thm:2-approx} are satisfied but when there are no precedence and time constraints. We now prove that the aforementioned policy is also optimal when there is a time constraint but there are no precedence constraints. 

\begin{theorem}\label{thm:healthiest_optimal_noprecedence_timeconstraint}
Let there be $N(\geq2)$ nodes such that there are no precedence constraints (i.e., $\mathcal{E}=\emptyset$). Suppose Assumption \ref{assum:no_partaway_step} holds and $T \in \mathbb{N} \cup \{\infty\}$. Then, the policy that targets the healthiest node at each time-step while respecting the time constraint is optimal. 
\end{theorem}
\begin{IEEEproof}
Consider any optimal (non-jumping) sequence $A= \{i_1,\ldots,i_x\}$ that permanently repairs $x(\le N)$ nodes. Let $T' (\le T+1)$ be the total number of time-steps taken by the sequence $A$ to permanently repair all $x$ nodes.\footnote{Note that the total number of time-steps is one larger than the largest time-step index $T$ because the first time-step in a sequence is time-step 0.} Then,
\begin{equation}
    T'=\frac{1-E_1}{\Delta_{inc}}+\frac{1-E_2}{\Delta_{inc}}+\ldots+\frac{1-E_x}{\Delta_{inc}}, \label{eq:time_expr}
\end{equation}
where $E_j$ is the health of node $i_j$ when it is reached in the sequence, i.e., $E_1=v_0^{i_1}$ and $E_k=v_0^{i_k} - n \sum_{j = 2}^{k} \left(1-E_{j-1}\right)$ for $ k \in \{2, \ldots, x\}$. Alternatively, $E_k=v_0^{i_1}n\left(1+n\right)^{k-2}+v^{i_2}_0n(1+n)^{k-3}+\ldots+v_0^{i_{k-1}}n+v^{i_k}_0-n\left(1+n\right)^{k-2}-n(1+n)^{k-3}-\ldots-n$ for all $ k \in \{2, \ldots, x\}$. Note that $E_1$ is the largest when node $i_1$ has the largest initial health; $E_2$ is the largest when node $i_1$ has the largest initial health and node $i_2$ has the second largest initial health (as the coefficients of $v_0^{i_1}$ and $v_0^{i_2}$ are $n$ and 1, respectively); $E_3$ is the largest when node $i_1$ has the largest initial health, node $i_2$ has the second largest initial health and node $i_3$ has the third largest initial health (as the coefficients of $v_0^{i_1}$, $v_0^{i_2}$ and $v_0^{i_3}$ are $n(1+n)$, $n$ and 1, respectively); and so on. Thus, the non-jumping sequence $B$ that targets the nodes in decreasing order of their initial health is optimal when time is constrained because it permanently repairs $x$ nodes in less time in comparison to the time taken by sequence $A$ to permanently repair $x$ nodes (as the value of $T'$ in \eqref{eq:time_expr} for sequence $B$ is less than the corresponding value for sequence $A$). Therefore, the policy that targets the healthiest node at each time-step while respecting the time constraint is optimal by Theorem \ref{thm:nonjumping_optimal_notallfixed}.  
\end{IEEEproof}

We also show that the policy of targeting the healthiest node at each time-step (while respecting the precedence and time constraints) is  optimal in certain classes of DAGs. We start with the following definition.
\begin{definition}
Given a DAG, the nodes that do not have any incoming edges are said to belong to the ``first level"; after removing all the nodes in the first level and all the outgoing edges of the nodes in the first level, the nodes that do not have any incoming edges are said to belong to the ``second level", and so on. A DAG in which all the nodes in a level have incoming edges from all the nodes in the previous level is termed as a ``complete series graph".
\end{definition}

\begin{figure}[ht]
	\begin{center}
		\includegraphics[scale=0.35]{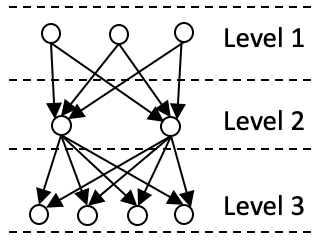}
	\end{center}	
	\caption{An example of a \textit{complete series} graph.}
	\label{series}
\end{figure}

Figure \ref{series} shows a \textit{complete series} graph that has three levels.   

\begin{proposition}\label{prop:completeseries_decreaselarger}
Let there be $N(\geq2)$ nodes with precedence constraints given by a \textit{complete series} graph $G=\{\mathcal{V},\mathcal{E}\}$. Suppose Assumption \ref{assum:no_partaway_step} holds and $T\in\mathbb{N}\cup \{\infty\}$. Then, the policy that targets the healthiest node at each time-step while respecting the precedence and time constraints is  optimal.
\end{proposition}

\begin{IEEEproof}
Due to precedence constraints, we need to permanently repair all the nodes in the first level before targeting the other nodes in a \textit{complete series} graph. Since the optimal policy is to target the healthiest node at each time-step (while respecting the time constraint) when there are no precedence constraints (by Theorem \ref{thm:healthiest_optimal_noprecedence_timeconstraint}), we follow this policy for the nodes in the first level. After permanently repairing all the nodes in the first level, we need to permanently repair all the nodes in the second level before targeting other nodes. Thus, the policy of targeting the healthiest node at each time-step (while respecting the time constraint) in the second level is followed after permanently repairing all the nodes in the first level. The rest of the proof follows the above argument.
\end{IEEEproof}

So far, we assumed that $\Delta_{dec}^j\ge \Delta_{inc}^j, \forall j\in \{1,\ldots,N\}$. In the next section, we analyze other cases.
\section{Control Sequences for $\Delta_{\MakeLowercase{dec}}^\MakeLowercase{j}< \Delta_{\MakeLowercase{inc}}^\MakeLowercase{j}, \quad \forall \MakeLowercase{j}\in \{1,\ldots,N\}$} \label{sec:repair_much_larger}
We now consider the case when repair rates are sufficiently larger than the deterioration rates, i.e., $\Delta_{inc}^j>(N-1)\Delta_{dec}^j, \forall j \in \{1,\ldots,N\}$ and $\Delta_{inc}^j >\sum_{l\neq j} \Delta_{dec}^l, \quad \forall j \in \{1,\ldots,N\}$. We will use the concept of a \textit{modified health value} defined as follows.

\begin{definition}
The modified health value of a node at a time-step is the health value of a node at that time-step minus its deterioration rate. 
\end{definition}
\subsection{Near-optimal policy}
Consider a directed rooted tree where the total number of nodes is at most equal to $k(\ge 1)$. In the context of disaster recovery, such a tree can represent damaged infrastructure components (with dependencies) within a particular neighborhood. We will consider such trees for the representation of precedence constraints as follows. 

\begin{theorem} \label{thm:ktree}
Let there be $N(\geq2)$ nodes with precedence constraints represented by a graph $G=\{\mathcal{V},\mathcal{E}\}$, which is a set of disjoint rooted trees that each contain at most $k(\ge 1)$ nodes. Suppose $\Delta_{inc}^j>(N-1)\Delta_{dec}^j, \forall j \in \{1,\ldots,N\}$, $\Delta_{inc}^j >\sum_{l\neq j} \Delta_{dec}^l, \quad \forall j \in \{1,\ldots,N\}$ and $T=\infty$. Then, the policy that targets the node with the least modified health value at each time-step while respecting the precedence constraints is $\frac{1}{k}$-optimal.
\end{theorem}

\begin{IEEEproof}
Let $\mathcal{R}$ be the set of root nodes of the trees. Let $A$ denote the sequence that targets the node with the least modified health value in the set $\mathcal{R}$ at each time-step, and let $B$ denote the optimal sequence. Let the number of nodes that are permanently repaired by sequences $A$ and $B$ be $y$ and $x$, respectively. We will show that $y\ge \frac{x}{k}$. We prove this by contradiction. Suppose $\frac{x}{k}>y$.  Denote the set of root nodes that are permanently repaired in sequence $B$ by $\mathcal{F} \subseteq \mathcal{R}$, and let $z \triangleq|\mathcal{F}|$. Note that $z\ge \left \lceil\frac{x}{k} \right \rceil$ because the root node in a tree should be permanently repaired before targeting other nodes in the tree. Then, $z \ge y+1$ because $\frac{x}{k}>y$ implies $\left \lceil\frac{x}{k} \right \rceil \ge y+1$. We now construct a sequence $C$ by modifying sequence $B$ such that at every time-step of sequence $B$ where a node not belonging to the set $\mathcal{F}$ is targeted, we replace that node with a node from the set $\mathcal{F}$. Then, sequence $C$ also permanently repairs all the nodes in the set $\mathcal{F}$ because the nodes of set $\mathcal{F}$ are targeted more frequently in sequence $C$ than in sequence $B$. Since the optimal policy is to target the node with the least modified health value at each time-step when there are no precedence and time  constraints (by Proposition 1 of \cite{hgehlot_tac}), we obtain a contradiction because sequence $C$ permanently repairs $z\ge y+1$ nodes whereas sequence $A$ (which is the optimal sequence to permanently repair the set of root nodes) permanently repairs $y$ nodes. Therefore, the assumption that $\frac{x}{k}>y$ is not true.

Let the sequence that targets the node with the least modified health value (in the set $\mathcal{V}$) at each time-step while respecting the precedence constraints be denoted by $D$. We now argue that sequence $D$ permanently repairs at least as many nodes as sequence $A$. Denote the set of root nodes that are permanently repaired in sequence $A$ by $\mathcal{Z}\subseteq \mathcal{R}$ and the number of nodes in the set $\mathcal{Z} $ by $y$. Then, there exists an ordering of nodes in the set $\mathcal{Z}$, denoted by $\{i_1,i_2,\ldots,i_y\}$, such that the initial health values of those nodes satisfy 
\begin{equation}
    v_0^{i_j}>(y-j)\Delta_{dec}^{i_j}, \quad \forall j \in \{1,\ldots,y\} \label{eq:nec_conditions}
\end{equation}
because of the following argument. Let $\mathcal{A}_t \subseteq \mathcal{Z}$ denote the set of nodes that have not been targeted at least once by the entity prior to time $t$ in sequence $A$. Note that $\mathcal{Z}=\mathcal{A}_0\supseteq \mathcal{A}_1\supseteq \ldots\supseteq \mathcal{A}_{y-1}$. At time $t=0$, $|\mathcal{A}_t|=y$, where $|\mathcal{A}_t| $ denotes the cardinality of set $\mathcal{A}_t$. At $t=1$, $|\mathcal{A}_t|=y-1$ as there are $y-1$ nodes of the set $\mathcal{Z}$ that have not been targeted at least once by the entity in sequence $A$ and thus each node $k\in \mathcal{A}_1$ should have initial health value larger than $\Delta_{dec}^k$ to survive until $t=1$. At $t=2$, $|\mathcal{A}_t|\ge y-2$ as there are at least $y-2$ nodes of the set $\mathcal{Z}$ that have not been targeted at least once by the entity in sequence $A$ and thus each node $k\in \mathcal{A}_2$ should have initial health value larger than $2\Delta_{dec}^k$. Repeating this argument for the next $y-3$ time-steps, we can argue that there must be a permutation $\{i_1,\ldots,i_y\}$ of nodes that satisfies the conditions \eqref{eq:nec_conditions} in order for $y$ nodes to be permanently repaired in sequence $A$. 

After this, it can be argued that sequence $D$ permanently repairs at least $y$ nodes by the same way as in Proposition 1 of \cite{hgehlot_tac}. Let the number of nodes that are permanently repaired by sequence $D$ be $y'$. Then, $y'\ge y$ and therefore $\frac{x}{y'}\le \frac{x}{y}\le k$. Thus, the policy that targets the node with the least modified health value at each time-step while respecting the precedence constraints is $\frac{1}{k}$-optimal.  
\end{IEEEproof}

We now show that the factor of $\frac{1}{k}$ in Theorem \ref{thm:ktree} is sharp, in that there exist problem instances where targeting the node with the least modified health value at each time-step repairs only $\frac{1}{k}$ times the nodes as an optimal policy. 
\begin{example} \label{exmp:counterexample_repairlarger}
Consider the graph in Figure \ref{fig:counterexample}, which is a set of disjoint rooted trees that each contain at most two nodes. Let the initial health values of the nodes 1, 2 and 3 be 0.01, 0.02 and 0.8, respectively, and $T=\infty$. The deterioration and repair rates are homogeneous across all the nodes and are equal to 0.1 and 0.25, respectively. The least healthy node that can be targeted in the first time-step is node 1 (since deterioration rates are homogeneous across all the nodes, targeting the node with the least modified health value is equivalent to targeting the node with least health value). However, if node 1 is targeted in the first time-step then node 2 fails by the time the entity reaches it and we cannot target any more nodes. However, if node 2 is first permanently repaired then node 3 can also be permanently repaired. Note that although the policy that targets the least healthy node at each time-step while respecting the precedence constraints is not optimal in this example, it is indeed $\frac{1}{k}$-optimal as proved in Theorem \ref{thm:ktree}.
\end{example}

We now also provide an example to show that the policy that is characterized in Theorem \ref{thm:ktree} need not be $\frac{1}{k}$-optimal when $T\in \mathbb{N}$. 
\begin{example} \label{exmp:increasegreater_timeconstraint}
Consider a graph consisting of two nodes without any precedence constraints (i.e., the graph is a union of disjoint trees that each contain one node). Let the initial health values of the nodes 1 and 2 be 0.01 and 0.11, respectively. The deterioration and repair rates are homogeneous across all the nodes and are equal to 0.1 and 0.11, respectively, and $T=10$. If the policy of targeting the node with the least health value at each time-step is followed then no node gets permanently repaired while respecting the time constraint. However, if the non-jumping policy that first targets node 2 is followed, then node 2 can be permanently repaired. Thus, the policy that is characterized in Theorem \ref{thm:ktree} is not a 1-optimal policy (or optimal policy) in this example.
\end{example}

Therefore, characterizing near-optimal policies for this case under a time constraint is an avenue for future research.
\subsection{Optimal sequencing}
The paper \cite{hgehlot_tac} proved that the policy that targets the node with the least modified health value at each time-step is optimal when there are no precedence and time constraints. Examples \ref{exmp:counterexample_repairlarger} and \ref{exmp:increasegreater_timeconstraint} showed that this need not be true when there is a precedence or time constraint. 
However, the policy of targeting the node with the least modified health value at each time-step is optimal for special cases such as when the precedence constraints are given by a \textit{complete series} graph (defined in the previous section). 
\begin{proposition}
Let there be $N(\geq2)$ nodes with precedence constraints given by a \textit{complete series} graph $G=\{\mathcal{V},\mathcal{E}\}$. Suppose $\Delta_{inc}^j>(N-1)\Delta_{dec}^j, \forall j \in \{1,\ldots,N\}$, $\Delta_{inc}^j >\sum_{l\neq j} \Delta_{dec}^l, \quad \forall j \in \{1,\ldots,N\}$ and $T=\infty$. The optimal policy is to target the node with the least modified health at each time-step while respecting the precedence constraints. 
\end{proposition}

This result can be proved similarly as Proposition \ref{prop:completeseries_decreaselarger} by using the fact that the policy of targeting the node with the least modified health at each step is optimal when there are no precedence and time constraints \cite{hgehlot_tac}.

\section{Conclusions}
We studied the problem of finding (near-) optimal control policies for targeting different components whose states are disrupted, when there exist  precedence constraints between the components. We characterized control policies depending on the relationship between the repair and deterioration rates. There are several options for future extensions to this work: introducing stochasticity in the health values, extending the problem to time-dependent deterioration and repair rates, characterizing policies with improved ratios $\alpha$, and characterizing near-optimal policies when for all $j\in \mathcal{V}$, $\Delta_{dec}^j<\Delta_{inc}^j<(N-1)\Delta_{dec}^j$. 

\section{Appendix}
In this section, we prove that Problem \ref{problem} is NP-hard. 
This proof is inspired from the paper \cite{garey1976scheduling}; however, the job scheduling problem in \cite{garey1976scheduling}  does not consider deterioration of jobs and there are additional differences between the problem that is considered in \cite{garey1976scheduling} and Problem \ref{problem} as mentioned in the review of job scheduling studies in Section \ref{sec:intro}. We start by defining the NP-complete \textit{Clique} problem \cite{cormen2009introduction} and an instance of a decision version of Problem \ref{problem} referred to as $OR_d$.   

\begin{problem}[\textit{Clique}]
Given an undirected graph $G'=(\mathcal{V'},\mathcal{E'})$ consisting of $s$ vertices and $q$ edges, and a positive integer $p(\le s)$, does $G'$ have a complete subgraph of size $p$, i.e., a set of $p$ vertices such that each pair of vertices in the set is connected by an edge in $\mathcal{E'}$?
\end{problem}
\begin{problem}[$OR_d$]
Given a directed acyclic graph $G=\{\mathcal{V},\mathcal{E}\}$ consisting of $N(\ge 2)$ nodes with initial health values $\{v_0^j\}$, along with repair and deterioration rates $\{\Delta_{inc}^j\}$ and $\{\Delta_{dec}^j\}$, respectively, $T=\infty$, and an integer $z$ such that $0\le z\le N$, is there a non-jumping control sequence $u^{\ast}_{0:T}$ that respects the precedence constraints and gives a reward $J(v_0, u^{\ast}_{0:T})\ge N-z$?
\end{problem}

We now present the proof of Theorem \ref{thm:NP-hard}.
 
\begin{IEEEproof}
    Given an instance of Clique we construct an instance of $OR_d$ as follows. We construct a total of $N=s+q+1$ nodes that are of three types as follows: a \textit{v}-node is constructed corresponding to each vertex in $G'$, an \textit{e}-node is constructed corresponding to each edge in $G'$, and a \textit{root} node $r$ is constructed. The parameters of these nodes are set as follows. For each \textit{v}-node $j$, set $v_0^j=\frac{2(2^{s+q}-1)+1}{2+2(2^{s+q}-1)+1}$ and $\Delta_{dec}^j=\Delta_{inc}^j=\frac{1}{2+2(2^{s+q}-1)+1}$. For each \textit{e}-node $j$, set $v_0^j=\frac{2(2^{\frac{p(p+1)}{2}}-1)+1}{2+2(2^{\frac{p(p+1)}{2}}-1)+1}$ and $\Delta_{dec}^j=\Delta_{inc}^j=\frac{1}{2+2(2^{\frac{p(p+1)}{2}}-1)+1}$. The root node $r$ has the same parameters as a \textit{v}-node. We construct directed edges such that there is an edge starting from a \textit{v}-node and ending in an \textit{e}-node if the vertex corresponding to the \textit{v}-node lies at one of the ends of the edge (in graph $G'$) corresponding to the \textit{e}-node. We also construct directed edges starting from the root node and ending in all the other nodes, so that a DAG is formed. Finally, set $z=q-\frac{p(p-1)}{2}$. Note that the constructed instance of $OR_d$ is polynomial in the size of the given instance of Clique. 

	We will now prove that the answer to the given instance of Clique is \textit{yes} if and only if the answer to the constructed instance of $OR_d$ is \textit{yes}. Suppose that the answer to the instance of Clique is \textit{yes}. Then, we will show that it is possible to create a non-jumping sequence for the constructed instance of $OR_d$ in which no more than $q-\frac{p(p-1)}{2}$ nodes permanently fail while respecting the precedence constraints. The first node in the created sequence is node $r$ because of the precedence constraints. Note that node $r$ is permanently repaired in the created sequence because the first node in any non-jumping sequence is always permanently repaired (when $T=\infty$). After permanently repairing node $r$, $p$ \textit{v}-nodes whose corresponding vertices form a complete graph in the Clique are targeted in the created sequence.
	Note that it takes two time-steps to permanently repair node $r$ in the created sequence because $\frac{1-v_0^r}{\Delta_{dec}^r}=2$. 
	Also, the health value of each \textit{v}-node $j$ after two time-steps is equal to $v_2^j=v_0^j-2\Delta_{dec}^j=\frac{2(2^{s+q}-1)+1-2}{2+2(2^{s+q}-1)+1}$. 
Thus, the number of time-steps it takes to permanently repair the first \textit{v}-node $j$ in the created sequence is equal to $\frac{1-v_2^j}{\Delta_{dec}^j}=4$. Thus, the second \textit{v}-node $j$ starts getting targeted after six time-steps in the created sequence and at that time its health value is equal to  $v_6^j=v_0^j-6\Delta_{dec}^j=\frac{2(2^{s+q}-1)+1-6}{2+2(2^{s+q}-1)+1}$. Thus, the number of time-steps it takes to permanently repair the second \textit{v}-node $j$ in the created sequence is equal to $\frac{1-v_6^j}{\Delta_{dec}^j}=8$.
	 Proceeding in this way, it can be shown that the total number of time-steps taken to permanently repair node $r$ and $p-1$ \textit{v}-nodes is equal to $2(2^{p}-1)$ because the $i$th node in the created sequence  (where $1\le i\le p$) takes $2^i$ time-steps to get permanently repaired. Note that $\frac{v_0^j}{\Delta_{dec}^j}=2\left(2^{s+q}-1\right)+1$ for all \textit{v}-nodes $j$, and represents the number of time-steps it takes for $j$ to permanently fail if it is not targeted within the first $\frac{v_0^j}{\Delta_{dec}^j}$ time-steps. Thus, for all \textit{v}-nodes $j$, we define $\gamma_{\textit{v}}=\frac{v_0^j}{\Delta_{dec}^j}=2(2^{s+q}-1)+1$. Note that the time-step at which the $p$th \textit{v}-node starts getting targeted in the created sequence is less than $\gamma_{\textit{v}}$ because $2(2^{p}-1)\le 2^{s+1}-2< 2^{s+1}-1\le 2(2^{s+q}-1)+1$ (as $p\le s$ and $q\ge 0$). Thus, all the $p$ \textit{v}-nodes are permanently repaired in the created sequence. After permanently repairing $p$ \textit{v}-nodes that correspond to the solution of Clique, $\frac{p(p-1)}{2}$ \textit{e}-nodes that correspond to the edges of the complete subgraph in the solution of Clique are targeted. Note that $\frac{v_0^j}{\Delta_{dec}^j}=2(2^{\frac{p(p+1)}{2}}-1)+1$ for all \textit{e}-nodes $j$. Thus, for all \textit{e}-nodes $j$, we define $\gamma_{\textit{e}}=\frac{v_0^j}{\Delta_{dec}^j}=2(2^{\frac{p(p+1)}{2}}-1)+1$. Note that there are $1+p+\frac{p(p-1)}{2}-1=\frac{p(p+1)}{2}$ nodes that are targeted before the last \textit{e}-node among the aforementioned \textit{e}-nodes in the created sequence. 
	 Thus, the last \textit{e}-node among the aforementioned \textit{e}-nodes starts getting targeted in the created sequence at time-step $2(2^{\frac{p(p+1)}{2}}-1)$ because the $i$th node in the created sequence  (where $1\le i\le \frac{p(p+1)}{2}$) takes $2^i$ time-steps to get permanently repaired.
	 Thus, all the aforementioned \textit{e}-nodes get permanently repaired because  $2(2^{\frac{p(p+1)}{2}}-1)<\gamma_{\textit{e}}$. 
	Finally, the remaining $s-p$ \textit{v}-nodes are targeted in the created sequence. Note that there are $1+p+\frac{p(p-1)}{2}+s-p-1=s+\frac{p(p-1)}{2}$ nodes that are targeted before the last \textit{v}-node in the created sequence. Thus, it takes $2(2^{s+\frac{p(p-1)}{2}}-1)$ time-steps in order to start targeting the last \textit{v}-node because the $i$th node in the created sequence (where $1\le i\le s+\frac{p(p-1)}{2}$) takes $2^i$ time-steps to get permanently repaired. Thus, 
	it is possible to permanently repair all the \textit{v}-nodes because $2(2^{s+\frac{p(p-1)}{2}}-1)<\gamma_{\textit{v}}$ (as $q\ge \frac{p(p-1)}{2}$). Thus, except the remaining $q-\frac{p(p-1)}{2}$ \textit{e}-nodes, all the nodes are permanently repaired in the created sequence.
    
	Now we show the opposite direction. Suppose the answer to the constructed instance of $OR_d$ is \textit{yes}. Then, there exists a non-jumping sequence in which at most $z=q-\frac{p(p-1)}{2}$ nodes permanently fail. Note that there are at most $N-1=s+q$ nodes that are permanently repaired before the last node that is targeted in the given sequence. Thus, the largest time-step at which the last targeted node in the given non-jumping sequence starts getting targeted is equal to $2(2^{s+q}-1)$ because the $i$th node in the given sequence takes $2^i$ time-steps to get permanently repaired. Thus, all the \textit{v}-nodes are permanently repaired in the given sequence because $2(2^{s+q}-1)< 2(2^{s+q}-1)+1=\gamma_{\textit{v}}$. Note that the first node that is targeted in the given  sequence is node $r$ because of the precedence constraints. Since the first node in the given sequence is permanently repaired and all \textit{v}-nodes are also permanently repaired, the set of all the nodes that permanently fail in the given sequence is a subset of all the \textit{e}-nodes. The number of \textit{e}-nodes that are permanently repaired in the given sequence is at least equal to $\frac{p(p-1)}{2}$ because at most $q-\frac{p(p-1)}{2}$ nodes permanently fail in the given sequence and the total number of \textit{e}-nodes is equal to $q$. 
	Therefore, at least $p$ \textit{v}-nodes need to be permanently repaired before at least $\frac{p(p-1)}{2}$ \textit{e}-nodes can be permanently repaired in the given sequence because of the following two reasons. First, a \textit{v}-node is an in-neighbor of an \textit{e}-node if the vertex corresponding to the \textit{v}-node lies at one of the ends of the edge (in graph $G'$) corresponding to the \textit{e}-node. Secondly, among all the undirected graphs that have $\frac{p(p-1)}{2}$ edges, a complete graph (containing $p$ vertices) has the least number of vertices. Note that all the \textit{e}-nodes that are permanently repaired in the given sequence start getting targeted before time-step $\gamma_{\textit{e}}=2(2^{\frac{p(p+1)}{2}}-1)+1$. Also, the maximum number of nodes (when the $i$th node takes $2^i$ time-steps to get permanently repaired) that can be targeted such that the last node starts getting targeted before time-step $2(2^{\frac{p(p+1)}{2}}-1)+1$ is equal to $\frac{p(p+1)}{2}+1$. Thus, node $r$, $p$ \textit{v}-nodes and $\frac{p(p-1)}{2}$ \textit{e}-nodes are targeted before time-step $2(2^{\frac{p(p+1)}{2}}-1)+1$ in the given sequence such that the vertices corresponding to the $p$ \textit{v}-nodes form a complete subgraph of size $p$ in graph $G'$. 
	
	Since non-jumping sequences are optimal when the precedence constraints are given by a DAG, $\Delta_{dec}^j \ge \Delta_{inc}^j, \forall j\in\{1,\ldots,N\}$ and $T\in \mathbb{N} \cup \{\infty\}$ by Theorem \ref{thm:nonjumping_optimal_notallfixed}, we conclude that Problem \ref{problem} is NP-hard.
\end{IEEEproof}

\bibliographystyle{IEEEtran} 
\bibliography{refs}

\begin{IEEEbiography}[{\includegraphics[width=1in,height=1.25in,clip,keepaspectratio]{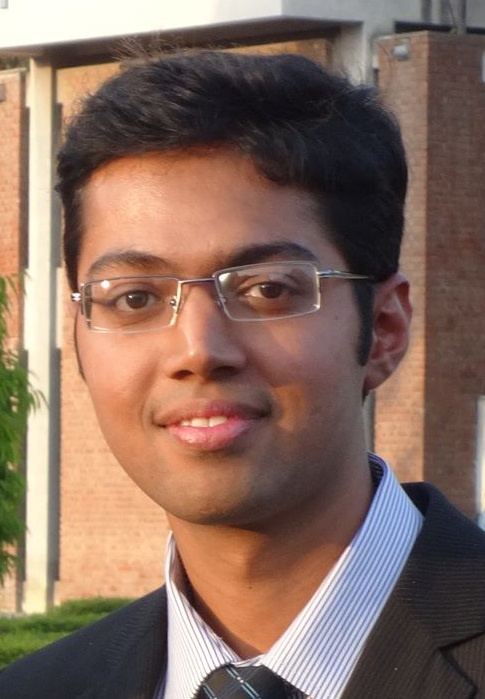}}]{Hemant~Gehlot}
is a PhD candidate in the Lyles School of Civil Engineering at Purdue University, being supervised by Dr. Satish V. Ukkusuri and Dr. Shreyas Sundaram. He received his BTech and MTech degrees from the Indian Institute of Technology Kanpur in 2015. He was a finalist for the Best Student Paper Award at the IFAC Workshop on Distributed Estimation and Control in Networked Systems (NecSys) 2019. His research interests include optimal control and combinatorial optimization. 
\end{IEEEbiography}

\begin{IEEEbiography}[{\includegraphics[width=1in,height=1.25in,clip,keepaspectratio]{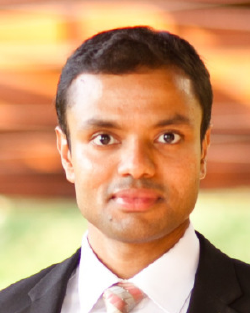}}]{Shreyas~Sundaram}
is an Associate Professor in the School of Electrical and Computer Engineering at Purdue University. He received his MS and PhD degrees in Electrical Engineering from the University of Illinois at Urbana-Champaign in 2005 and 2009, respectively. He was a Postdoctoral Researcher at the University of Pennsylvania from 2009 to 2010, and an Assistant Professor in the Department of Electrical and Computer Engineering at the University of Waterloo from 2010 to 2014. He is a recipient of the NSF CAREER award, and an Air Force Research Lab Summer Faculty Fellowship. At Purdue, he received the Hesselberth Award for Teaching Excellence and the Ruth and Joel Spira Outstanding Teacher Award. At Waterloo, he received the Department of Electrical and Computer Engineering Research Award and the Faculty of Engineering Distinguished Performance Award. He received the M. E. Van Valkenburg Graduate Research Award and the Robert T. Chien Memorial Award from the University of Illinois, and he was a finalist for the Best Student Paper Award at the 2007 and 2008 American Control Conferences. His research interests include network science, analysis of large-scale dynamical systems, fault-tolerant and secure control, linear system and estimation theory, game theory, and the application of algebraic graph theory to system analysis.
\end{IEEEbiography}

\begin{IEEEbiography}[{\includegraphics[width=1in,height=1.25in,clip,keepaspectratio]{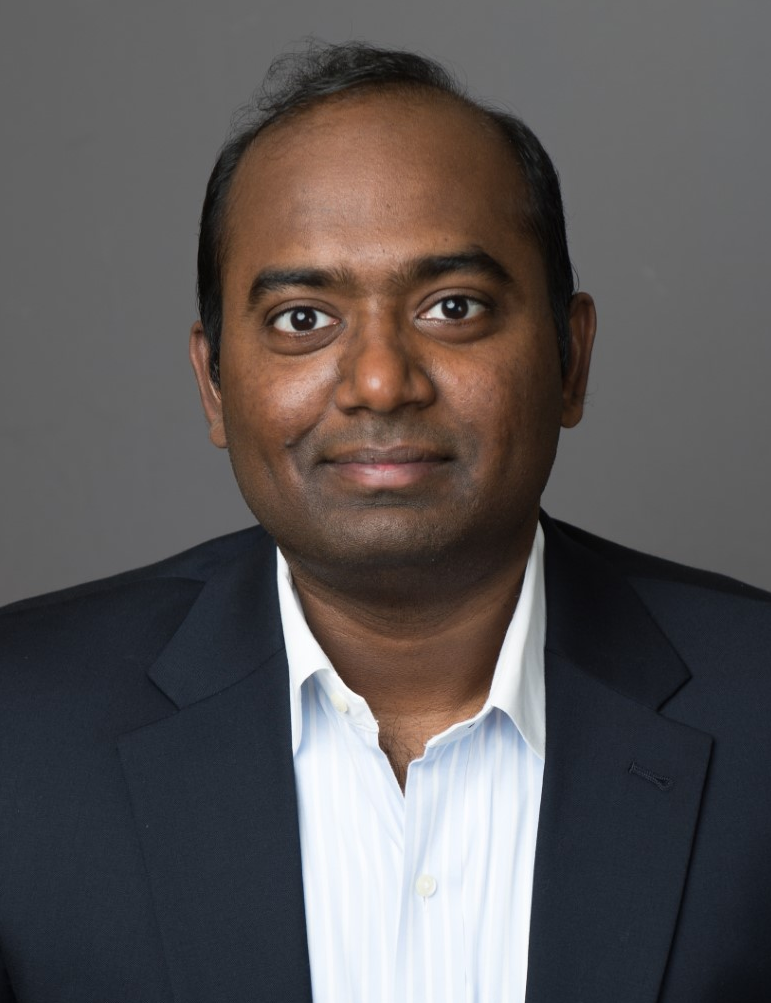}}]{Satish~V.~Ukkusuri}
is a Professor in the Lyles School of Civil Engineering and Director of the Urban Mobility Networks and Intelligence Lab at Purdue University. His research is in the area of interdisciplinary transportation networks with current interests in data driven mobility solutions, disaster management, resilience of interdependent networks, connected and autonomous traffic systems, dynamic traffic networks and smart logistics.  He has published more than 350 peer reviewed journal and conference articles on these topics.
\end{IEEEbiography}
\end{document}